\documentclass[preprint,showpacs,showkeys,preprintnumbers,amsmath,amssymb]{revtex4}

\usepackage{epsfig}
\usepackage{graphicx}
\usepackage{dcolumn}
\usepackage{bm}

\begin{document}

\preprint{Review Copy}

\title{A Non-linear Generalization of Singular Value Decomposition \\ and its Application to
Cryptanalysis}
\author{Prabhakar G. Vaidya.}
\author{Sajini Anand P. S.}%
 \email{sajini@nias.iisc.ernet.in}
\author{Nithin Nagaraj.}%
\affiliation{%
School of Natural and Engineering Sciences, National Institute of
Advanced Studies, IISc Campus, Bangalore 12.}

\homepage{http://www.iisc.ernet.in/nias}

\date{\today}

\begin{abstract}

    Singular Value Decomposition (SVD) is a powerful tool in linear
algebra and has been extensively applied to Signal Processing,
Statistical Analysis and Mathematical Modeling. We propose an
extension of SVD for both the qualitative detection and quantitative
determination of nonlinearity in a time series. The paper
illustrates nonlinear SVD with the help of  data generated from
nonlinear maps and flows (differential equations). The method is to
augment the embedding matrix with additional nonlinear columns
derived from the initial embedding vectors and extract the nonlinear
relationship using SVD. The paper also demonstrates an application
of nonlinear SVD to cryptanalysis where the encrypted signal is
generated by a nonlinear transformation. A comparison of the method
for both noise-free and noisy data along with their surrogate
counterparts is included.

\end{abstract}


\maketitle

\textbf{Singular Value Decomposition (SVD) along with its related
variation known as Principal Component Analysis is a powerful
technique for data analysis in linear algebra which has found lot of
applications in various fields such as Signal Processing,
Statistical Analysis, Biomedical Engineering , Genetics Analysis,
Mathematical and Statistical Models, Graph Theory, Psychology etc
\cite{Intro}. In this paper, we discuss an extension of SVD for both
the qualitative detection and quantitative determination of
nonlinearity in a time series. SVD is performed on the embedding
matrix created from data series. The conventional SVD can determine
the form of linear relationships among data vectors. For the
proposed method, the embedding matrix is augmented by nonlinear
columns derived from the usual ones. Now if the SVD gives zero
singular values there is a linear relationship present among the
columns. In that case, we could exactly determine the nonlinearity
present in the data. The paper also demonstrates an application of
nonlinear SVD to cryptanalysis where the encrypted signal is
generated by a nonlinear transformation. Nonlinear methods are
useful in cryptography as the signals to be decrypted are often
generated by non-linear transformations. We have included examples
of maps (Logistic map and Henon map) and flows (Van der Pol
oscillator and Duffing oscillator) to illustrate the method of
nonlinear SVD to identify parameters. The paper presents the
recovery of parameters in the following scenarios: (i) data
generated by maps and flows (ii) Comparison of the method for both
noisy and noise-free data (iii) Surrogate data Analysis for both the
noisy and noise-free cases}

\section{\label{sec:intro} Introduction}

Historically SVD has been used for finding the dimension of a linear
system as it gives statistically independent set of variables which
could span the state space. Standard SVD based methods known as
Singular Spectrum Analysis were used for detecting nonlinearity in a
qualitative manner ~\cite{broomhead}. An abrupt decrease in the
profile of the singular spectrum is an indication of lower
dimensional determinism. But this method fails to distinguish
between a chaotic data and its surrogates ~\cite{Sahay, Abarbanel}.
The failure of this method for some well known chaotic processes,
and when the data is corrupted with noise are also reported
~\cite{Kugiumtzis, Buzug}. Bhattacharya and kanjilal  proposed a
method of quadratic scaling of singular values in order to detect
the determinism in time series \cite{Bhattacharya}. The decreasing
singular values were weighted more to highlight the deterministic
and stochastic features. This method could qualitatively distinguish
between the data series and its surrogates.

The Grassberger- Procaccia (G - P) algorithm was used to show that a
finite correlation dimension of an irregular time series  is an
indication of underlying deterministic nonlinearity
\cite{Grassberger}. G - P algorithm has a few drawbacks as it  fails
when the data is noisy and is unable to distinguish stochastic
processes with power-law power-spectra from chaos \cite{Osborne,
Ruelle}. There are various methods to detect the underlying
determinism in a time series. One uses the distribution of
correlation coefficients as a qualitative method to distinguish
chaos from noise because the spectrum is flat for noise but
gradually decays for chaos~\cite{Sugihara}. But this method fails to
distinguish between correlated noise and chaos \cite{Cazellus,
Ensminger}. Similarly in Statistics, the distribution of sample
autocorrelation function (ACF) is used as an important tool to
assess the degree of dependance in data in order to select a model
~\cite{statistics text book}. A constant sample ACF, which takes
zero values for all the lags (delays) is an indication of data being
independent and identically distributed (iid) noise. If the sample
ACF spectrum shows decay or oscillations appropriate statistical
parametric models could be used to model the data. There are
developments in the field of  both linear and nonlinear versions of
Autoregressive(AR) and Autoregressive Moving Average(ARMA) models
along with many robust algorithms; e. g. the Fast Orthogonal Search
method which can obtain correct model parameters irrespective of the
model selection ~\cite{Korenberg}.

If one is interested in a detection of the existence of nonlinearity
but not the determination of the underlying model A. Porta et. al.
suggest an alternate method ~\cite{porta1,porta2}. Their method
consists of using Takens embedding of data in a higher dimensional
phase space and then subdividing the phase space into
non-overlapping hypercubes. Prediction is based on the behavior of
the median member of each hypercube. They find that an error
function dips much further with actual data than with its
surrogates.

Linear and nonlinear AR and ARMA models are efficiently used for
determining the nonlinearity; i.e. to find the parameters of the
appropriate models using optimal parameter search (OPS) algorithms
~\cite{Sheng Lu} and through various least square techniques: Least
Square~\cite{strang}, Total Least Square~\cite{TLS} and Minimizing
the Hypersurface Distance~\cite{Sheng Lu2}. In this paper, we
discuss a nonlinear extension of SVD to detect and exactly determine
the nonlinearity with  an application to cryptography. The proposed
method of nonlinear SVD for maps is similar to the nonlinear AR and
ARMA regression proposed by Lu. S et al~\cite{Sheng Lu} and
Marmarelis ~\cite{Marmarelis}. But the method is not limited to
polynomial regression. Any deterministic nonlinearity present in the
data could possibly be recovered. Nonlinear SVD method could be
considered as functional regression in an extended phase space.

Chaotic cryptanalysis is an emerging field of chaotic cryptography
which deals with the breaking of secret codes without any access to
the super keys or parameters of the system ~\cite{PGV,PGV-a,PGV-b}.
It often deals with the problem of system identification from the
encrypted data which could be noisy and incomplete. Consider two
parties communicating across a private channel. The aim of
cryptanalysis is to decrypt the message. The signal sent across the
channel looks random but it is generated by a deterministic
dynamical system. Since the intended recipient has some information
about the system parameters (a key) she can retrieve the information
from the encrypted signal. All that the cryptanalyst knows about the
signal is that it must have been generated by a deterministic
dynamical system. He has no clue about the parameters or the
dimension of the system. Here we are proposing a new cryptanalysis
tool based on SVD to find the information about a system from the
encrypted signal. Nonlinear SVD and time delayed embedding together
with the method of finding derivatives from data ~\cite{PGV1} can be
used to identify the nonlinearity of the system from the time
series.

The paper is organized as follows. Section I is introduction.
Section II briefly reviews the  conventional SVD technique. Section
III describes the method of nonlinear SVD and Section IV discuss the
applicability of the method to data series. A numerical example of
retrieval of Logistic map parameter from data is discussed in
Section V. Section VI gives the comparison of nonlinear SVD and
conventional SVD analysis of the data and its surrogates. Section
VII discusses the extension of the method for higher order maps.
Section VIII shows the numerical results in the presence of two
types of noises: Uniform noise and Gaussian noise. Section IX is the
extension of the method to flows and recovery of the nonlinearity
from chaotic data is explained in Section X.  Section XI is a
discussion on linear and nonlinear models. Section XII is discussion
on cryptanalysis and  the conclusions is Section XIII.

\section{\label{sec:level 1} Singular Value Decomposition}

Singular Value Decomposition can be considered as a generalization
of the spectral-decomposition of square matrices, to analyze
rectangular matrices. SVD decomposes a rectangular matrix into three
simple matrices: two orthogonal matrices and one diagonal matrix
~\cite{strang}. In general, SVD theorem can be stated as follows:
any $m\times n$ matrix $A$, with $m \geq n$ can be factored into
three matrices: $U$ (column orthogonal, $m \times n$ matrix), $W$
(diagonal,$n \times n$ matrix) and $V$ (orthogonal $n \times n$
matrix). When $A$ is real, $A = U.W.V^T$ (where $V^T$ is the
transpose of $V$). For complex matrices, $W$ remains real but $U$
and $V$ become unitary. 
The diagonal elements of $W$ matrix are known as the singular values
of $A$.

This decomposition is a technique that works well with matrices that
are either singular or else numerically very close to singular. SVD
is also used to calculate pseudo-inverses when the natural inverse
of the matrix does not exist ~\cite{N.Recipe}. SVD and
pseudo-inverses are generally used in statistics for solving least
square problems. Data compression using SVD is one of the standard
applications in image processing ~\cite{strang,kalman}.

\section{\label{sec:level 1} Method of Nonlinear Singular Value Decomposition}

Given a time series generated by any system $\{ X \}= \{X_{1},
X_{2}, X_{3},\ldots, X_{N} \}$ the aim of nonlinear system
identification is (i) to detect if there is nonlinearity and (ii) to
exactly determine the equation which generated the data. The current
study is based on the data generated by nonlinear maps and flows. We
will begin by using the standard Takens embedding: a method of
reconstruction of the state space with time delayed data segments
known as embedding vectors ~\cite{Takens}. The embedding matrix $E$
is created from the time delayed vectors as follows. A typical
embedding vector $Y^i$ is the $m$ dimensional embedding vector
generated from the given time series.
\begin{eqnarray*}
Y^1 & = &  (X_{1} \ X_{2} \ldots X_{m} \ )^T .\\
Y^2 & = & (X_{2} \ X_{3} \ldots X_{m+1} \ )^T.\\
& \ldots &   \\
Y^i & = & (X_{i} \ X_{i+1} \ldots  X_{m+i} \ )^T.
\end{eqnarray*}
Note that $(X_{1} \ X_{2} \ldots X_{m} \ )^T$ is a column vector and
denotes the transpose of $(X_{1} \ X_{2} \ldots X_{m} \ )$.  The
collection $ \{Y^i \}$ is the time delay embedding of the given
data. Let the embedding matrix $E$ be created from $k$ embedding
vectors as follows.
\begin{equation}
E = \left[\begin{small}
 \begin{array}
{cccccc}
& (Y^1)^T \\
& (Y^2)^T \\
& \ldots \\
& \ldots\\
& \\
& (Y^{k})^T \\
 \end{array}\end{small}
\right]= \left[\begin{small}
 \begin{array}{cccccc}
 X_{1} & X_{2}&  & \ldots &  &  X_{m}  \\
 X_{2} & X_{3}&  & \ldots &  &  X_{m+1} \\
   &   &   & \ldots &  &  \\
X_{k} & X_{k+1}&  & \ldots &  &  X_{m+k} \\
 \end{array}\end{small}
\right]. \end{equation}

For nonlinear SVD, we extend the embedding matrix $E$ by adding
nonlinear columns. Let F be the extended embedding matrix.
\begin{equation}
F = [E \ :  \ f_1 \ f_2 \ \ldots f_i \ ]
\end{equation}
The last $i$ columns of $F$ matrix, $f_1, f_2 \ldots f_i$ are
functions of the columns of $E$ matrix, $ \{ E^{<1>}, E^{<2>} \ldots
E^{<m>} \}$ where $ E^{<i>}$ denotes the $i^{th}$ column of $E$. In
general, a non-linear column refers to the square, cube, any other
higher powers of a column, the product of two or more columns or any
other kind of non-linearity such as exponentiation and trigonometric
functions of the column. If there is a non-linear relationship
between the time delayed vectors, it could be interpreted as a
linear relationship between the time-delayed vectors and
corresponding nonlinear columns. The dimension of $F$ matrix is $k
\times p$ where $ p = (m+i)$. This extended embedding matrix can be
considered as a higher dimensional linear system. The singular value
decomposition can find the linear relation between the embedding
vectors and corresponding nonlinear columns, thus recovering the
nonlinear relationship inherent in the data. SVD is performed on F
to get,
\begin{equation}
 \ \ F   = U W V^T.
\end{equation}
Therefore,
\begin{equation}
F V = U W.
\end{equation}
Expanding Eq. 4
\begin{equation}
F.\left[\begin{small}
 \begin{array}{cccccc}
 V^{<1>} & V^{<2>}& \ldots &  V^{<p>}\\
\end{array}\end{small}
\right]= \left[\begin{small}
 \begin{array}{cccccc}
 U^{<1>} & U^{<2>} & \ldots & U^{<p>}\\
\end{array}\end{small}
\right]. \left[\begin{small}
 \begin{array}{cccccc}
 W_{1, 1} & 0   & 0   &\ldots  &   0 \\
 0  & W_{2, 2}   &  0  & \ldots &   0 \\
 0  & 0  &   & \ldots   & \ldots \\
  0 &   \ldots &   & \ldots   & W_{p, p} \\
 \end{array}\end{small}
\right] \end{equation} Using partitions of $V$ and $W$ and expanding
along the last column of $V$ and $W$ we get,
\begin{equation}
\left[\begin{small}
 \begin{array}{cccccc}
 F^{<1>} & F^{<2>}   & \ldots  & F^{<p>}\\
\end{array}\end{small}
\right].\left[\begin{small}
 \begin{array}{cccccc}
& V_{1, p} \\
& V_{2, p}\\
& \ldots\\
& V_{p, p} \\
\end{array}\end{small}
\right]=\left[\begin{small}
 \begin{array}{cccccc}
 U^{<1>} & U^{<2>}  &\ldots  & U^{<p>}\\
\end{array}\end{small}
\right]. \left[\begin{small}
 \begin{array}
{cccccc}
& 0\\
& 0\\
& \ldots\\
& W_{p, p} \\
 \end{array}\end{small}
\right].\end{equation}
\begin{equation}
V_{1, p}(F^{<1>}) + V_{2, p}(F^{<2>}) + \ldots + V_{p, p}(F^{<p>}) =
0 (U^{<1>})+ 0 (U^{<2>})+ \ldots + W_{p,p}(U^{<p>})
\end{equation}(Note that $U^{<i>}$ stands for the $i^{th}$ column of $U$ matrix.
$W_{i,j}$ is the element on $i^{th}$ row and $j^{th}$ column of $W$
matrix). If the $p^{th}$ singular value of W is zero, $$W_{p, p} =
0.$$ then
\begin{equation}
V_{1, p} (F^{<1>}) + V_{2, p} (F^{<2>})+ \ldots + V_{p, p} (F^{<p>})
= 0.
\end{equation}
Or, in another notation,
\begin{equation}
\sum_{p}{F_{j,p}. V_{p,n}}= 0 \ \ \ \forall \ j
\end{equation}

Eq. 8 and 9 can be seen as a statement that the columns of $F$
including those formed by the nonlinear functions $\{ f_1, f_2
\ldots f_i \}$ now span a linear vector space. This equation is true
for all rows of $F$ and hence by exploiting this relation, the
dependance between the linear and nonlinear columns of the $F$
matrix can be recovered.

\section{\label{sec:level 1} Nonlinear SVD of data}

We begin with our assumption that the underlying equation is a
function of the delay vectors. In case the data is noise free and
the conventional SVD of data does not result in a small enough
singular value then we try different $F$'s of the form $F= [E \ : \
f_1 \ f_2 \ \ldots f_i \ ]$ as shown in Section III. We keep trying
different $f_i$'s till the nonlinear SVD gives at least one nearly
zero singular value. Next section shows a numerical example of data
generated by Logistic map. We made a simple guess for the nonlinear
function and it worked for that case. If it did not,we would have
tried other functions.

But when the data is noisy, the singular value $W_{p, p} \neq 0 $
even if we get the right $F$ matrix of nonlinear functions.  Hence
we try different nonlinear functions and choose the $F$ that gives
the lowest value of $W_{p, p}$. Ideally we want $\frac { W_{p,p} }
{W_{1,1}}$ (the ratio of $p^{th}$ singular value to the $1^{st}$
singular value) to be below some preset criterion.  As the noise
level in the data increases chances are higher that the method of
nonlinear SVD fails. Therefore our confidence in the estimated model
equation goes down with the increase in noise.

\section{\label{sec:level 1} Numerical Example}

Consider the data generated by a Logistic map $X_{n+1} =
\lambda.X_{n}.(1-X_{n})$ where $ 0 \leq X <1 $ and $\lambda$ is an
unknown parameter. We show in this section how $\lambda$ could be
retrieved from the data. Let the data series sampled at a chosen
time delay $\tau$  be $\{X_{n}\}= {\{X_{1}, X_{2},\ldots, X_{N}}\}$.
The data can be embedded in three dimensions using embedding vectors
as shown in the following matrix :
\begin{equation*}
E = \left[\begin{small}
\begin{array}{cccccc}
X_1  & X_2   & X_3 \\
X_2  & X_3  & X_4  \\
& \ldots  & \\
X_{5}  & X_{6}  & X_{7}.
\end{array}\end{small}
\right]. \end{equation*} The dimension of $E$ matrix is $5 \times
3$. After SVD operation on the embedding matrix $E$, we observed
that none of the singular values of  $E$ go to zero indicating no
linear dependance present in the data.  Now $E$ matrix is extended
to $F$ matrix as follows,
\begin{eqnarray*}
 F & = & [E \ :  \ f_1 \ ] \\
 &=& [ E^{<1>} \ E^{<2>} \ E^{<3>} \ f_1 \ ]\\
 &=& [ E^{<1>} \ E^{<2>} \ E^{<3>} \ {E^{<1>}}^2 \ ]
\end{eqnarray*}
So that a typical row of $F$ is seen as $[\ X_p \ X_{p+1} \ X_{p+2}\
{X_p}^2 \ ]$ where $p$ goes from $1$ to $4$. SVD of $F$ gives a zero
singular value $W_{4, 4} = 0$ indicating a linear dependence between
the first column $X_n$ and the added nonlinear column ${X_n}^2$. Eq.
8 for this particular numerical example is,
\begin{equation}
V_{1,4}(X_n)+ V_{2,4} (X_{n+1}) + V_{3,4} (X_{n+2})+ V_{4,4}
(X_{n}^2)= 0.
\end{equation}

Substituting the numerical values, the exact equation can be
recovered from Eq. 9 as follows.
\begin{eqnarray*}
-0.696311 X_n  + 0.174078 X_{n+1}+ 0 X_{n+2}  + 0.696311 X_{n}^2  & = & 0 .\\
 4X_n -X_{n+1} - 4{X_{n}}^2  & = & 0. \\
 4{X_n}(1-{X_{n}}) & = & X_{n+1}.
\end{eqnarray*}
For the numerical example, the initial condition that generated the
trajectory was $0.02$ and the selected parameter value $\lambda$ of
the logistic map was $4$. We have observed that the method of
nonlinear SVD works reasonably well even in the presence of noise,
provided the noise is below some threshold value. Let the data
$\{X_{n}\}$ be contaminated by additive noise $\{P_{n}\}$ which is
either gaussian or uniform, $ \{ \widehat{X}_n \} = \{X_{n}\} +
\{P_{n}\}.$ The same procedure can be done on the embedding matrix
$F$ created from the noisy data $\{{\widehat{X}_n}\}$ to extract the
nonlinearity. TABLE. 1 shows the estimated values of parameter
$\lambda$ for the data generated by Logistic family of maps under
the presence of different types of noises. The preset criterion for
the noisy case was $10^{-6}$. For recovering the parameters, we made
an assumption that the singular values smaller than this can be
considered zero and the underlying equation is extracted as
explained in the noise-free case.


\section{\label{sec:level 1} Comparison of Non-linear SVD with standard SVD and Surrogates}

Fig. 1 shows the quadratically scaled singular value spectrum:
$n^2.\sigma_n$ versus $n$; where $\sigma_n$ is the $n^{th}$ singular
value generated by the standard SVD on the embedding matrix $E$. The
time series $\{X\}$ generated from the Logistic map: $X_{n+1} =
\lambda X_n(1-X_n)$ where $0 \leq X_n \leq 1$ and $\lambda = 4$ is
used to create an embedding matrix $E$ with unity delay as explained
in section III. The dimension of the embedding matrix is selected as
$21 \times 21$ for this particular example. Standard SVD operation
gives 21 non-zero singular values: Note that the profile gradually
increases and slowly comes down. Similarly for the nonlinear SVD
operation, the embedding matrix $F$ is generated as explained in
section IV. The dimension of the embedding matrix is kept same as
that of $E$ i.e. $21 \times 21$ of which the last 10 columns are
squares of the first 10 columns. Now SVD operation gives 21 singular
values, out of which the last 10 are zero. Observe that the
nonlinear SVD profile is significantly different from that of
conventional SVD case as the former is `flat' towards the end. Fig.
1, the spectrum for nonlinear SVD shows zero singular values
compared to the standard SVD. There is a significant qualitative
change in the spectrum. The profile of nonlinear SVD case drops to
zero rapidly compared to the standard SVD case.

\begin{figure}
\centering
\includegraphics[scale=1]{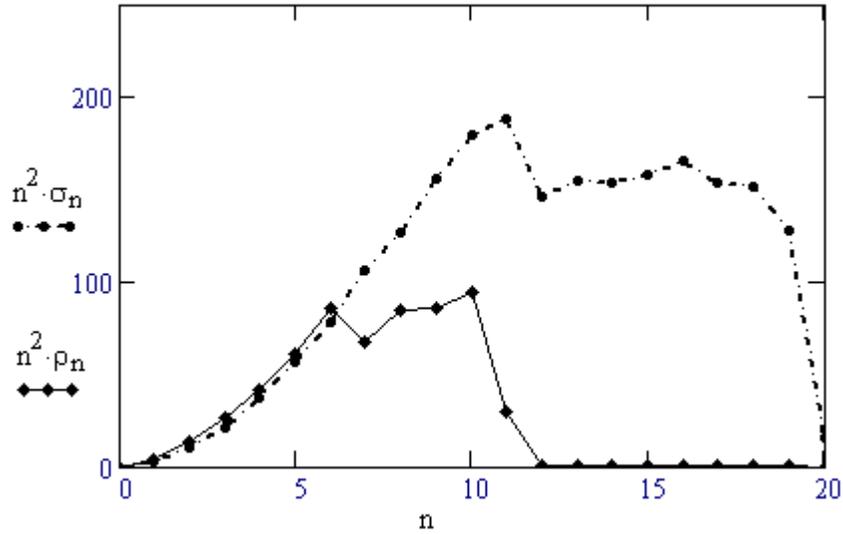}
\caption{\label{spectrum} quadratically scaled spectrum of singular
values for the case of standard SVD $(n^2.\sigma_n)$ and non-linear
SVD $(n^2.\rho_n)$. }
\end{figure}

Similar analysis is done for the surrogates. Surrogates are
generated from the Fourier Transform of the data by randomizing the
phases. K surrogate data series $\{Xsurr(k)\}$ are generated from
$\{X\}$ such that $\{Xsurr(k)\}$ and $\{X\}$ have the same power
spectrum ~\cite{Theiler Kennel}. Hence $\{Xsurr(k)\}$ is considered
as the nondeterministic counterpart of $\{X \}$ . Fig. 2 (i) and
(ii) show the quadratically scaled spectra of singular values of the
data and its surrogates for the standard SVD and nonlinear SVD
respectively under noise free conditions. We observe that the
selection of nonlinear columns has worked since the nonlinear SVD
has given identically zero singular values as shown (ii) of Fig. 2.
Moreover the method is able to clearly distinguish between the data
and the surrogates. The figures also contain the spectrum for the
original data $\{X \}$ for a comparison with its surrogates. Similar
analysis for noisy data is included in section VIII.

\begin{figure}
\centering
\includegraphics[scale=0.7]{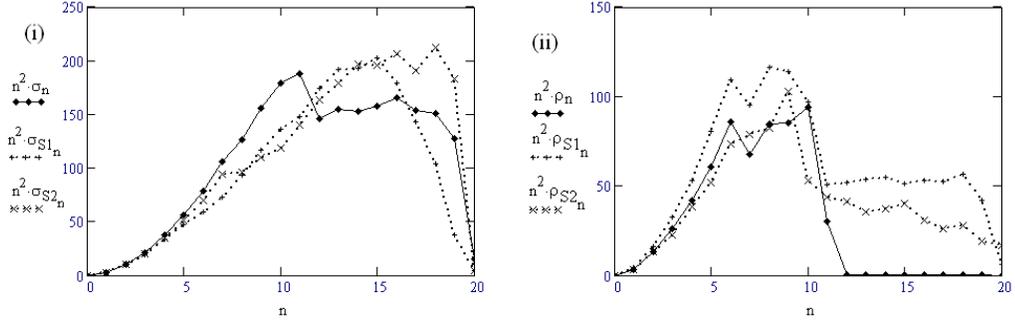}
\caption{\label{spectrum} (i) Quadratically scaled spectrum of
singular values by standard SVD on the data $(n^2.\sigma_n)$ and its
surrogates $(n^2.\sigma_{S1_n})$ and $(n^2.\sigma_{S2_n})$. (ii)
Similar spectrum by nonlinear SVD for the data $(n^2.\rho_n)$ and
its surrogates $(n^2.\rho_{S1_n})$ and $(n^2.\rho_{S2_n})$. }
\end{figure}

Assume that we have varied the number and types of the nonlinear
terms present in the embedding matrix $F$. For the case of data from
quadratic map, we have added cubic columns in the F matrix instead
of quadratic columns. The singular value spectrum looks
qualitatively the similar to the quadratic case as shown in Fig. 3.
But the exact relationship cannot be retrieved quantitatively, as
none of the singular values go to zero. We need to try different
embedding matrices and select the one which gives at least one
nearly zero singular value.

\begin{figure}
\centering
\includegraphics[scale=1]{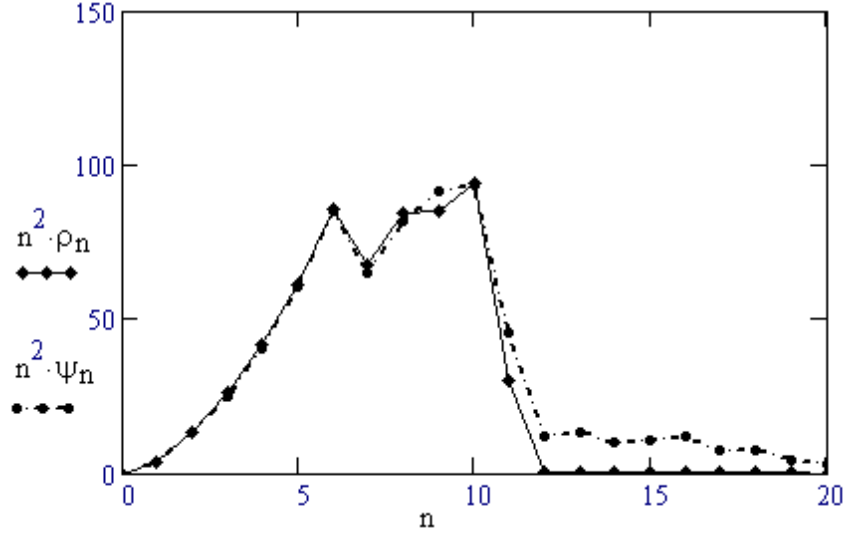}
\caption{\label{spectrum}  quadratically scaled singular value
spectra by nonlinear SVD for different choices of $F$ matrices on
data generated by Logistic map(i) $n^2.\rho_n$ versus $n$ ; where
$\rho_n$ is the $n^{th}$ singular value of $F$ with quadratic
columns, and (ii) $n^2.\phi_n$ versus $n$ ; where $\phi_n$ is the
$n^{th}$ singular value of $F$ with cubic columns }
\end{figure}

\section{\label{sec:level} Recovering non-linearity:  Higher order maps}

Let us now discuss the problem of recovering the non-linear equation
when the data is generated from a higher order map of the following
form,
\begin{equation}
X_{n+2} = f( X_n, X_{n+1}).
\end{equation}
The well known Henon map, falls in this category.
\begin{eqnarray}
X_{n+1} &=& c - a X_{n}^2 + Y_n.  \\
Y_{n+1} &=& b X_{n}.
\end{eqnarray}
We generated some data using this map and later assumed that only
the $X$ data is available. In that case it is more convenient to
list this in the form of Eq. 11. With that form in mind we set up F
to be ${[\ 1 \ X_{n+2} \ X_{n+1} \ X_n \ {X_n}^2 \ X_{n+1}^2 \
(X_{n+1}X_n) \ ]}$.  Performing SVD on the embedding matrix $F$ we
get the following relationship between the iterates.
\begin{eqnarray*}
0 &=& 0.496904 - 0.6956656 X_{n+1}^2 + 0.1490712 X_{n} - 0.496904
{X_{n+2}}.\\
0 &= &1 - 1.4 X_{n+1}^2 + 0.3 X_{n} - {X_{n+2}}.
\end{eqnarray*}
Therefore,
\begin{equation}
{X_{n+2}} = 1 - 1.4 X_{n+1}^2 + 0.3 X_{n}.
\end{equation}
This can be seen as equivalent to Eq. 12 and 13 with parameter
values $a = 1.4 \ b = 0.3$ and $c = 1$. Thus using the proposed
method the parameters are retrieved from the X data.

Similarly Consider the case of the Logistic map again. We are once
again required to find the parameters $\lambda$, but all the od
iterates of the time series are suppressed. In this case we could
consider the map of the form,
\begin{equation}
X_{n+2} = g( X_n).
\end{equation}

The non-linear SVD on the modified data can retrieve a quartic
nonlinearity. The predicted equation, for this particular example is
of the form $g(x) = B x + C x^2 + D x^3 + E x^4$ where $B, C, D, E$
are functions of the parameter $\lambda$.  SVD on $F=[\ X_n \
X_{n+2} \ {X_n}^2 \ {X_n}^3 \ {X_n}^4] $ gives a zero singular value
corresponding to the following relationship between the columns.
\begin{equation*}
0 = 14.7609 X_n - X_{n+2}- 71.4725 {X_n}^2 + 113.4232 {X_n}^3 -
56.7116 {X_n}^4.
\end{equation*}
Therefore,
\begin{equation}
X_{n+2} = 14.7609 X_n - 71.4725 {X_n}^2 + 113.4232 {X_n}^3 - 56.7116
{X_n}^4.
\end{equation}
But for the Logistic map $X_{n+1} = \lambda X_n(1-X_n)$. Hence the
second iterate can be written as a function of its first iterate as
follows,
\begin{equation}
X_{n+2} = (\lambda^2) X_n - (\lambda^2 + \lambda^3) {X_n}^2 + (2
\lambda^3) {X_n}^3 - (\lambda ^3) {X_n}^4.
\end{equation}
Comparing Eq. $16$ and $17$ we recover the value of the parameter
$\lambda= 3.842$.

\section{\label{sec:level} Numerical results for noisy data}

\begin{figure}
\begin{center}
\includegraphics[scale=0.8]{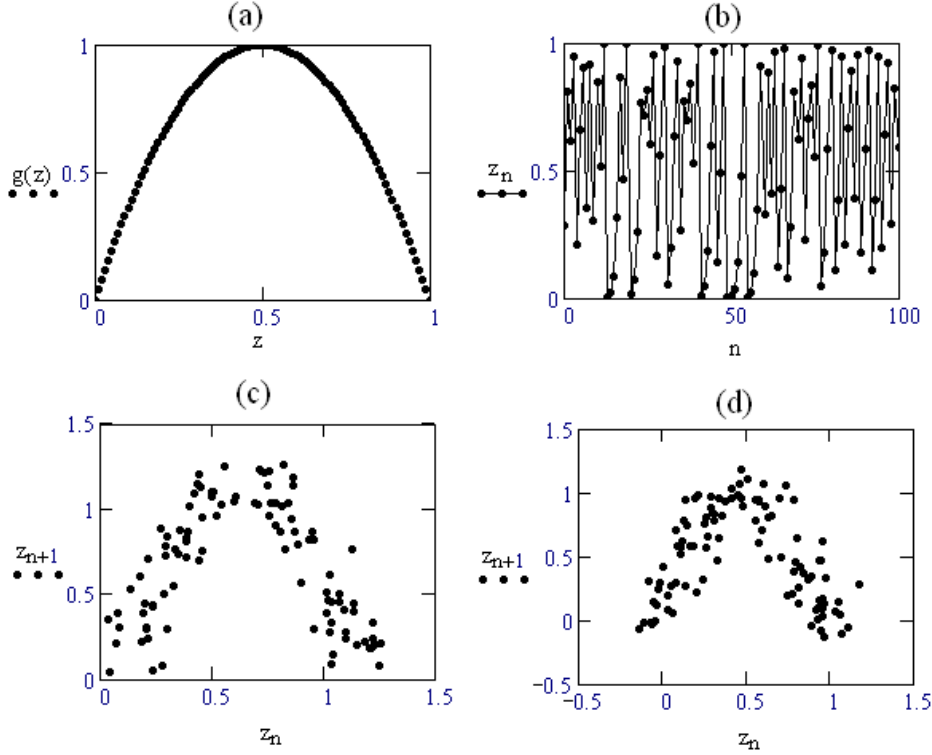}
\caption{\label{ss}Time series (a) the chaotic data generated from
the Logistic map: $X \mapsto 4X(1-X)$ where  $0 \leq X \leq 1$ (b)
The phase space. (c) Phase space for noisy data (added uniform noise
N(0,1)with noise level 28.089 \%) (d) Phase space for noisy data
(added gaussian noise N(0,1)with noise level 28.123 \%). \emph{Noise
level is defined as the ratio of the maximum noise value to the
maximum signal value.} }
\end{center}
\end{figure}

Fig. 3 (c) and (d) show the state space created from the noisy data
for both the uniform and gaussian noises. Conventional SVD on the
noisy data gives the value of $W_{p, p}/ W_{1,1} $ somewhere in the
range $(10^ {-1},10^ {-3})$ for different noise levels of the data
from Logistic and Henon Maps shown in TABLE I and II. For nonlinear
SVD, we have set value of the criterion  $\frac { W_{p,p} }
{W_{1,1}}$ as $ 10^ {-6}$. If the ratio $\frac { W_{p,p} }
{W_{1,1}}$ is below the preset value $10^ {-6}$ the parameters are
retrieved from the data. As the noise increases the method almost
breaks down and even fails to distinguish between the data and its
surrogates.

We found that when the noise level was low, the non-linear SVD was
able to recover the non-linearity. TABLE. 1 shows the estimated
values of parameter $\hat{a}, \hat{b} $ for the data generated by
Logistic family of maps: $X_{n+1}= a_1 X_n - a_2 {X_n}^2$ where $0
\leq X_n \leq 1$ for the values of $a = 4$ and $b = 4$ for different
Peak Signal to Noise Ratio (PSNR) using the nonlinear SVD algorithm.
PSNR is defined as $20.\log \{ \max(\textrm{Signal})/ \sqrt{MSE}\}$
where MSE is the Mean Squared Error, the average of the square of
the noise added to the signal. The proposed method works well when
the PSNR is above $28$ for Gaussian noise and $33$ for Uniform
noise. It seems that the method breaks down as the noise content in
the signal increases (i.e. for lower PSNR values). Similarly Table.
II shows the estimated values for parameter $\hat{a}, \hat{b},
\hat{c}$ for the Henon map ${X_{n+1}= c -a X_{n}^2 + Y_n ; \ Y_{n+1}
= b X_{n}}$ in the presence of noise using nonlinear SVD. Again the
method breaks down for higher noise contents (PSNR $< 30$) as we saw
in the Logistic case.

\begin{table}[!h]
\centering \caption{Estimated values of estimates $\hat{a_1},
\hat{a_2} $ for the Logistic map: $X_{n+1}= a_1 X_n - a_2 {X_n}^2$
where $0 \leq X_n \leq 1$ for the parameter values $a_1 = 4$ and
$a_2 = 4$ in the presence of noise using nonlinear SVD. Peak Signal
to Noise Ratio (PSNR) defined as $20.\log \{ \max(\textrm{Signal})/
\sqrt{MSE}\}$.}\vspace{0.1in}
\begin{tabular}{|c|c|c|c|c|c|c|}

\hline Parameters & \multicolumn{3}{c|}{Noise (Uniform
Distribution)}& \multicolumn{3}{c|}{Noise (Gaussian
Distribution)}\\\cline{2-7} & $\hat{a_1}$ &
$\hat{a_2}$ &  PSNR & $\hat{a_1}$ & $\hat{a_2}$ &  PSNR \\
\hline
             & 4.0367 & 3.9941 & 44.2314 & 4.099 & 4.0879 &43.4502 \\
$a_1$ = 4    & 4.1427 & 4.0206 & 38.7308  & 4.106& 4.0919 & 41.8398\\
$a_2$ = 4    & 3.8757 & 3.8473 & 37.7963  & 4.247& 4.1568 &37.095\\
             & 4.2358 & 4.1516 & 36.4626 & 4.299& 4.1484 & 34.5231\\
            & 4.3895 & 4.2506  & 33.6095  & 4.1563& 3.9952 & 32.8561\\
            & 4.4809 & 4.2783 & 30.7117  & 4.0301& 3.824 & 28.8170\\
             & 4.429 & 3.8507 & 24.6316 & 3.8041& 2.5627 & 24.151 \\
\hline
\end{tabular}
\end{table}

\begin{table}[!h]
\centering \caption{Estimated values of parameters $\hat{a},
\hat{b}, \hat{c} $ for the Henon map ${X_{n+1}= c -a X_{n}^2 + Y_n ;
\ Y_{n+1} = b X_{n}}$ in the presence of noise using nonlinear SVD.
}\vspace{0.1in}
\begin{tabular}{|c|c|c|c|c|c|c|c|c|}

\hline Parameters & \multicolumn{4}{c|}{Noise (Uniform
Distribution)}&\multicolumn{4}{c|}{Noise (Gaussian
Distribution)}\\\cline{2-9} & $\hat{a}$ &  $\hat{b}$ & $\hat{c}$ &
PSNR & $\hat{a}$ &  $\hat{b}$& $\hat{c}$ & PSNR \\
\hline
         & 1.389 & 0.2896 & 1.003 & 46.863 & 1.4001 & 0.2998 & 1.003 &  44.5158\\
a = 1.4  & 1.3812 & 0.2923& 1.010 & 40.764 & 1.3983 & 0.2964 & 1.005 &   40.93\\
b = 0.3  & 1.357 & 0.2769  & 1.009 & 36.178 &1.4& 0.3167 & 0.9881 &   35.93\\
c = 1.0  & 1.3717 & 0.262  & 1.0204 & 32.975&1.4376& 0.3258 & 1.008 &  34.113\\
         & 1.3665 & 0.2758 & 1.067 & 30.4932 &  1.396& 0.3279& 1.04 &  31.336\\
         & 1.2969 & 0.2680 & 1.0349 & 26.549 & 1.5206& 0.3401 & 1.08 & 28.3245\\
         & 1.3268 & 0.1566 & 1.1489 & 20.8353 & 1.2512& 0.2917 & 0.9547 &  22.714\\
\hline
\end{tabular}
\end{table}

Fig. 4 shows the quadratically scaled singular spectra by standard
SVD on the logistic data $(n^2.\sigma_n)$ and its surrogates
$(n^2.\sigma_{S1_n})$ and $(n^2.\sigma_{S2_n})$ along with similar
spectrum by nonlinear SVD for the data $(n^2.\rho_n)$ and its
surrogates $(n^2.\rho_{S1_n})$ and $(n^2.\rho_{S2_n})$. Fig. 4 (i)
and (ii) show the case of gaussian noise N(0,1)  with noise level
$28.123 \%$ in the data. Fig. 4 (iii) and (iv) show noise added from
uniform distribution $(0,1)$ with noise level $28.089 \%$ . Noise
level is defined as the ratio of the maximum noise value to the
maximum signal value. The size of the embedding matrix is kept
constant for all the data and surrogate matrices for both the
standard and nonlinear SVD operation. The surrogate data spectrum is
added for the comparative analysis. It is clear that nonlinear SVD
distinguishes the original data from its surrogates even under the
presence of noise, provided the noise level is low.

\begin{figure}
\begin{center}
\includegraphics[scale=0.6]{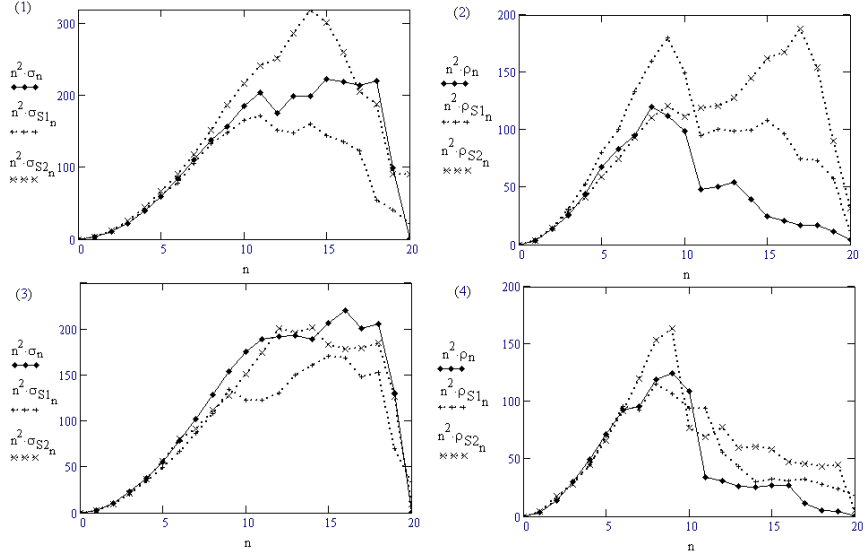}
 \caption{\label{Noise} Quadratically scaled spectrum of
singular values by standard SVD on the Logistic data
$(n^2.\sigma_n)$ and its surrogates $(n^2.\sigma_{S1_n})$ and
$(n^2.\sigma_{S2_n})$ along with similar spectrum by nonlinear SVD
for the data $(n^2.\rho_n)$ and its surrogates $(n^2.\rho_{S1_n})$
and $(n^2.\rho_{S2_n})$ (i) and (ii) for the case of gaussian noise
N(0,1) added noise level $27.484 \%$ (iii) and (iv) and for the case
of uniform noise [0,1) added noise level $28.182 \%$.}
\end{center}
\end{figure}

As we discussed in section I, if the goal is to detect the
nonlinearity but not to determine the underlying model, one could
use the method suggested by A. Porta et. al. In ref~\cite{porta1}
the case of the logistic model itself is discussed with parameter
value $\lambda =3.7$ under various noise levels. They have shown
that an error function dips much further with actual data than with
its surrogates. A comparison of this method and various other
methods are discussed in ref ~\cite{porta2}.

\section{\label{sec:level} Generalization to Differential Equations}

The same technique can be applied if the data is generated by a set
of nonlinear differential equations. Here the goal is to identify a
specific differential equation from the time series based on two
assumptions (i) the sampling is frequent (ii) noise level is very
low. We demonstrate here how this could be done by a combination of
nonlinear SVD and the method of finding accurate derivative from
data ~\cite{PGV1}. Consider the following narrative. Two teams A and
B are using a communication channel for sending information across
each other. The sender, A is generating data from a differential
equation of the form
\begin{eqnarray}
\frac{d}{dt}X_1 & = & X_2.\\
\frac{d}{dt} X_2 & =&  G( X_1, X_2).
\end{eqnarray}
A time series of length 50,000 was generated using Runge - Kutta
method from an initial condition $(1,-0.5)$ with sampling step size
0.001; from which a portion of $X_1$ of length 1000 was taken and
send to team B. A hint has been given that $G$ is a relatively
simple multinomial function.

\begin{figure}
\begin{center}
\includegraphics[scale=0.7]{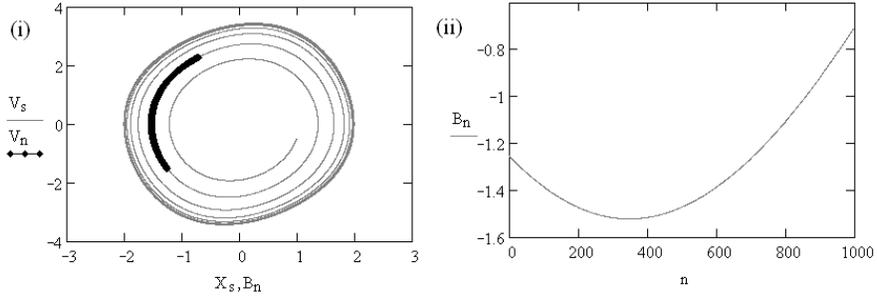}
\caption{\label{ss} (i) Phase space of the original data (${X,V}$)
generated by Team A ; the highlighted portion shows the section of a
short data segment of length 1000; B  which was sent to team B. (ii)
The temporal signal B.}
\end{center}
\end{figure}

We illustrate the method  and the issues involved by showing how
team B works for the unknown differential equation with a short time
series and the sampling interval. Team B is left with a small time
series $Y(t) = X_1 $ (accurate up to $10^{-16}$) shown in Fig. 5 (b)
along with an information that the step size was 0.001. B calculates
n derivatives of $Y(t)$ (${Y_2, Y_3,\ldots Y_{n+1}}$) from $Y(t)$ as
mentioned in ~\cite{PGV1}. $Y_n$ is the ${n-1}^{th}$ derivative of
$Y(t)$. $Y(t)$ is denoted as ${Y_1}$ \{the `$0^{th}$' derivative of
$Y(t)$\} for the rest of the section. As the first step B assumes
that $G$ is a nonlinear function consisting of linear and quadratic
terms of $Y_1$ and $Y_2$ as shown below. $$G = c_{1} Y_1 + c_{2} Y_2
+ c_{3} Y_1^2 + c_{4} Y_1.Y_2+ c_{5} Y_2^2.$$ Hence the columns of
the embedding matrix were chosen as $[{Y_3 \ Y_1 \ Y_2 \  {Y_1}^2 \
{Y_2}^2 \ (Y_1 Y_2) ]}$ and SVD is performed to get the singular
values $(201.33, \ 162.24, \ 83.95,\ 3.95,\ 0.44,\ 0.03)$. Since the
noise level is quite low, B decides to go to the next step of adding
cubic terms to the embedding matrix. Now the assumed nonlinear
function is, $$G = c_{1} Y_1 + c_{2} Y_2 + c_{3} Y_1^2 + c_{4}
Y_2^2+ c_{5}Y_1.Y_2  + c_{6} {Y_1}^3 + c_{7}{Y_2}^3 + c_{8}
({Y_1}^2. Y_2)+ c_{9} ({Y_2}^2.Y_1)]$$ and the corresponding
embedding matrix is
$${F = [{Y_3 \ Y_1 \ Y_2 \ {Y_1}^2 \ {Y_2}^2 \ (Y_1 Y_2) \ {Y_1}^3 \
{Y_2}^3 \ ({Y_1}^2. Y_2) \ ({Y_2}^2.Y_1) ]}}.$$ Singular values of
$F$ are (181.0635, \ 115.3297, \ 44.2941, \ 2.1510, \ 22.1215, \
0.1768, \ 0.0247, \ 0.0044, \ 0.0003, \ (2.995 $\cdot 10^{-11})$ ).
Since the last singular value is very small (in the order of
$10^{-11}$ ) B assumes it to be zero and tries to recover the linear
relationship corresponding to that singular value. Hence the
retrieved coefficient array is,
$$ \hat{C} = [ \ 1, \ 2.895, \ -0.237, \ 0, \ 0, \ 0, \ 0, \ 0, \ 0.237, \ 0 \
]^T.$$

Now team B has recovered the following equation,
\begin{eqnarray*}
0 &=& Y_3- 0.237 Y_2+ 0.237 Y_2 Y_1^2+ 2.895 Y_1.\\
Y_3&=& 0.237 Y_2- 0.237 Y_2 Y_1^2- 2.895 Y_1.\\
Y_3&=& 0.237 Y_2 (1- Y_1^2) - 2.895 Y_1.
\end{eqnarray*}
$Y_2$ and $Y_3$ are the $1^{st}$ and $2^{nd}$ derivatives of $Y_1$.
B has calculated 10 derivatives for this particular numerical
example. When Teams A and B get together B finds that A has used Van
der Pol equation of the following form with parameter values
$k=2.895$ and $c=0.237$.
\begin{eqnarray*}
\frac{d}{dt} X_1 &=& X_2.\\
\frac{d}{dt} X_2 &=& c X_2 (1- X_1^2) - k X_1.
\end{eqnarray*}
Team B' s estimated values are $k=2.895$ and $c=0.237$ which are in
agreement with the values used by A. The form of the equation and
the parameter values are exactly predicted by B using the proposed
method. Fig. 5 shows the phase space of Van der Pol oscillator and
the temporal signal which was sent to Team B.

\section{\label{sec:level} Chaotic data generated by a Differential Equations}

The same technique works for the chaotic data generated by a set of
nonlinear differential equations. Here we explain how to identify
the specific differential equation along with the parameters with
the help of nonlinear SVD and the method of finding accurate
derivative from data ~\cite{PGV1} as explained in previous section.
Consider the data generated by from the duffing equation of the
form.
\begin{eqnarray}
\frac{d}{dt}X_1 &=& X_2. \\
\frac{d}{dt} X_2 &=& -k X_1 - c X_2 - \delta {X_1}^2 + A.cos(\omega
t).
\end{eqnarray}
Let the parameter values be: $k=0.01$, $c=0.04496$, $\delta =1$,
$A=1.02$ and $\omega = 0.44964$. Parameters are selected such that
the system exhibit chaos. We ensured adherence to the conditions (i)
the sampling was frequent and (ii) noise level was very low. The
time series of short length $Y(t) = X_1 $ was sent to the receiver.
The temporal signal $Y(t)$ and phase space of the duffing oscillator
is displayed in Fig. 6. The receiver is informed that the data is
generated using an equation of the form,
\begin{eqnarray}
\frac{d}{dt}X_1 &=& X_2. \\
\frac{d}{dt} X_2 &=& G ( X_1, X_2)+ A.cos(\omega t).
\end{eqnarray}
Where $G$ is a low order polynomial function. Two additional
information (i) sampling step size $h = 0.01$ and (ii) forcing
frequency $\omega = 0.44964$ were given to the receiver. The
parameters and the exact form of the equation can be retrieved from
the data as shown below.
\begin{figure}
\begin{center}
\includegraphics[scale=0.6]{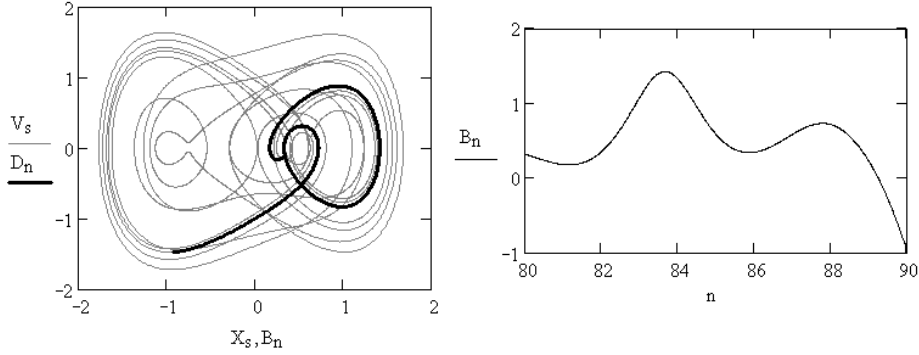}
\caption{\label{ss}  Phase space of the original data (${X,V}$)
generated the duffing oscillator under chaos ; the highlighted
portion shows the section of a short data segment of length 1000
which was used to identify the system.The temporal signal B is shown
on the right side.}
\end{center}
\end{figure}

Once again n derivatives (${Y_2, Y_3,\ldots Y_{n+1}}$) are
calculated. Now we assume that $G$ is a nonlinear function
consisting of linear, quadratic and cubic terms of $Y_1$ and $Y_2$
along with the sinusoidal functions. $G = c_{10} Y_1 + c_{01} Y_2 +
c_{20} Y_1^2 + c_{11} Y_1.Y_2+ c_{02} Y_2^2 + c_{30} {Y_1}^3 +
c_{03}{Y_2}^3 + c_{21} ({Y_1}^2. Y_2)+ c_{12} ({Y_2}^2.Y_1)]+
P.sin(\omega t) + Q.cos(\omega t)$. The corresponding embedding
matrix is  $F= [{Y_3 \ Y_1 \ Y_2 \ {Y_1}^2 \ {Y_2}^2 \ (Y_1 Y_2) \
{Y_1}^3 \ {Y_2}^3 \ ({Y_1}^2. Y_2) \ ({Y_2}^2.Y_1)\sin(\omega t)\
cos(\omega t)]}$. SVD operation on $F$ gives the set of singular
values (46.1294, \ 28.0900, \ 20.9784, \ 18.8628, \ 12.5940,\
6.8899, \ 4.8714, \ 4.2835, \ 2.0742, \ 0.8881, \ 0.5682, \ 0).
Since the last singular value is zero we can recover the linear
relationship corresponding to that singular value. The coefficient
array $\hat{C}$ is,
$$ \hat{C} = [ \ 1, \ 0.01, \ 0.04496, \ 0, \ 0, \ 0, \ 1, \ 0, \ 0, \ 0,
\ -0.98958, \ 0.24723 ]^T.$$ The following equation is retrieved
from $\hat{C}$.
\begin{eqnarray*}
Y_3+ 0.01 Y_1 + 0.04496 Y_2 + 1. Y_1^3-
\sqrt{0.98958^2+0.24723}.cos(\omega t) &=& 0.\\
Y_3+ 0.01 Y_1 + 0.04496 Y_2 + 1. Y_1^3- 1.02 cos(\omega t) &=& 0.
\end{eqnarray*}
The estimated parameter values and the form of the equation are in
exact agreement with the Duffing equation from which the data was
generated.

\section{\label{sec:level} Linear versus nonlinear models}

A second order linear system $X_n = a.X_{n-1} + b.X_{n-2}$ was
simulated with parameter values $a=-1.5$ and $b=-1$. In the absence
of noise the linear SVD had a sharper fall off than the nonlinear
SVD. In the case of surrogate data neither linear SVD nor nonlinear
SVD had a sharp fall off. In such cases on the grounds of parsimony
alone the linear model would be chosen. In the presence of small
amount of noise the same situation continues. However when the noise
is beyond a certain value neither linear nor nonlinear SVD will show
a substantial qualitative difference with the surrogate data to have
any degree of confidence in either of the models

\section{\label{sec:level} Application to Cryptanalysis}

We could be interested in cryptanalysis ~\cite{crptography}, the
study of code breaking, which involves decrypting the encrypted
data, without an access to the secret key.  Suppose Alice and Bob
are communicating a secret across a private channel. The assumption
is that the data can be read only by Bob, the intended recipient,
and he has the key to decrypt the message.  The aim of cryptanalysis
is to decrypt the message without the key. The signal sent across
the channel looks random but it is generated by a deterministic
dynamical system. Since Bob has some information about the system
parameters (secret key) he can retrieve the information from the
encrypted signal. But the cryptanalyst, Eve is left with a random
signal which is to be decrypted.  All she knows about the signal is
that it must have been generated by a deterministic dynamical
system, preferably a non-linear one. She has no clue about the
parameters or the dimension of the system. The possibility of trial
and error is extremely tedious and time-consuming, and has a low
probability of being successful. As compared to code-making,
relatively little work has been done in cryptanalysis (hardly 1 in
100 papers propose a new method of cryptanalysis). Here we are
proposing a new method based on SVD to find the information about
the non-linear system from the encrypted signal. Nonlinear Singular
Value Decomposition and time delayed embedding can be used to
identify the nonlinearity of the system from the data.  As we have
discussed before, in cryptography the signals to be decoded are
mostly generated by non-linear systems and complete data is not
always available. A method which can extract the nonlinearity from
the encrypted signal could be useful for applications in
cryptanalysis. This method is potentially useful for chaotic
cryptography ~\cite{lakshmanan}.

\section{\label{sec:level} Conclusion}

We have proposed a non-linear extension of the singular value
decomposition (SVD) technique by means of appending additional
columns to the trajectory matrix which are non-linearly derived from
the existing columns. We propose nonlinear SVD as a method which is
useful for the qualitative detection and quantitative determination
of nonlinearity from a short time series. We have demonstrated the
utility of non-linear SVD for recovering the non-linear relationship
from time series generated by discrete and continuous dynamical
systems. As an example, we have demonstrated the results for the
data from Logistic map, Henon Map, Van der Pol Oscillator and
Duffing oscillator. The proposed method works quiet well in the
presence of noise, for both Gaussian and Uniform noise (provided the
noise level is not high). In principle, the method can work with any
type of non-linearity. The paper contains a comparative analysis of
the results for the data and its surrogates.

\begin{acknowledgments}
Authors thank Department of Science and Technology for supporting
the Ph.D programme at National Institute of Advanced Studies,
Bangalore.

\end{acknowledgments}

\end{document}